\documentclass[12pt]{article}
\usepackage{graphicx}
\usepackage{amssymb}
\usepackage{amscd}
\usepackage{amsmath}
\usepackage{cite}

\begin{document}
\begin{titlepage}

{\hbox to\hsize{\hfill July 2017 }}

\bigskip \vspace{3\baselineskip}

\begin{center}
{\bf \large 
Scale Invariant Top Condensate}

\bigskip

\bigskip

{\bf Archil Kobakhidze and Shelley Liang \\ }

\smallskip

{ \small \it
ARC Centre of Excellence for Particle Physics at the Terascale, \\
School of Physics, The University of Sydney, NSW 2006, Australia 
\\}

\bigskip
 
\bigskip

\bigskip

{\large \bf Abstract}

\end{center}
\noindent 
We propose a scale invariant model of dynamical
electroweak symmetry breaking via top condensation within the minimal 
top condensate see-saw model. The classical
scale invariance is realized nonlinearly by introducing conformal
compensator scalar field, the dilaton, which plays crucial role in
successful predictions for the Higgs boson. The
dilaton mass and its mixing with the composite Higgs boson appears
only at 2-loop level, both being proportional to the hierarchy between the electroweak and compositeness scales, $\propto v/\Lambda$. The mass scales in our model are generated via the mechanism of dimensional transmutation and the hierarchy $v/\Lambda \ll 1$ is radiatively stable, thus resolving the fine-tuning problem of the standard top condensate models. 
 \end{titlepage}

\section{Introduction}

The discovery of the Higgs boson completes the Standard Model (SM)
and confirms of the basic picture of mass generation through the spontaneous
electroweak symmetry breaking. However, the quadratic sensitivity
of the Higgs mass under the quantum correction from ultraviolet physics
and the related mass hierarchy problem remains a mystery. One attractive scenario for solving the hierarchy problem is the dynamical electroweak symmetry breaking. It is driven by a composite Higgs boson, that emerges as a bound state of more fundamental fermionic constituents due to some hypothetical strong attractive forces at $\Lambda \sim $ TeV scale \cite{Weinberg:1975gm, Susskind:1978ms}. There is no slightest experimental evidence in favour of this scenario so far. Current LHC data put severe constraints on the compositeness scale \cite{Arbey:2015exa}, somewhat undermining the validity of the Higgs compositeness hypothesis. 

One of the most economic models for dynamical electroweak symmetry breaking is the top condensate model \cite{Miransky:1988xi, Bardeen:1989ds, Cvetic:1997eb}. In this model the Higgs boson is an emergent low-energy degree of freedom of top and anti-top quarks bound state. In the minimal model, there is a correlation between the Higgs mass $m_h$ and top quark mass $m_t$, $m_h\approx 2m_t$ (in the top-loop approximation). In addition, for a large compositeness scale $\Lambda$ top quark mass is predicted (as opposed to being a free parameter within the Standard Model) through the infrared renormalisation group quasi-fixed point of the top-Yukawa coupling \cite{Pendleton:1980as}. Hence the top condensate provides more predictive framework for the electroweak symmetry breaking than the minimal Standard Model does. Unfortunately, the predicted top quark mass is larger than the experimentally measured one. This problem can be resolved in many extended models, perhaps the most economic being the so-called top condensate see-saw model (TCSM) \cite{Dobrescu:1997nm}. Nevertheless, the Higgs mass in many of the extended top-condensate models is predicted to be heavier than the top quark, which is in odds with observations. Furthermore, in models with large compositeness scale the solution of the mass hierarchy problem is compromised. These significant problems made many researchers to abandon the top condensate models.           

In a number of previous works we were advocating classical scale invariance as a potential symmetry that could be responsible for the stable mass hierarchies \cite{Foot:2007as}. In the conceptual framework an overall mass scale emerges  in a clasically scale invariant (massless) theory as a result of quantum fluctuations through the mechanism of dimensional transmutaion \cite{Coleman:1973jx}. This dimensional transmutation proliferates into the low energy sectors of the theory through feeble interactions and  results in technically stable mass hierarchies between high energy and low energy sectors. This scenario can be realised also within nonrenormalisable effective field theory models by introducing dilaton field. An explicit construction of the minimal Standard Model with hidden (non-linearly realised) scale invariance has been recently proposed in \cite{Kobakhidze:2017eml}.     

In this paper  we consider a scale invariant version of the minimal TCSM, where the standard model field content is complimented  by extra vector-like quarks and the dilaton. The dilaton originates from spontaneously broken scale invariance at high energies and is realised in the effective low-energy theory as a (pseudo)Goldstone boson. We find that experimentally observed values for the top quark and Higgs boson masses can be accommodated within this framework. Furthermore, the hierarchy of scales is stable under the radiative corrections owing to the underlying scale invariance. A rather generic prediction of this class of models is the
existence of a very light dilaton, which has interesting phenomenological
implications.

The paper is organised as follows. In the next section we introduce
the minimal scale invariant TCSM and derive some basic
predictions. In sec. 3 we briefly discuss possible constraints on
light dilaton and sec. 4 is reserved for conclusions.

\section{The minimal scale invariant TCSM}

The top quark condensate  see-saw model for dynamical electroweak symmetry breaking is an extension of the minimal top condensate model which, in addition to the third generation left-handed quark doublet, $Q_L=(t_L,b_L)^T$, and the right-handed top quark, $t_R$, involves a vector-like electroweak singlet quarks $\xi_R$ and $\xi_L$. The quantum numbers under the $SU(3)_C\times SU(2)_W\times U(1)_Y$ standard model symmetry group are as follows:   
\begin{equation}
Q_L:(\mathbf{3}, \mathbf{2},+1/3),~~~t_R, \xi_R, \xi_L:(\mathbf{3},\mathbf{1},+4/3)~.
\label{1}
\end{equation}
We assume that some unspecified strong dynamics\footnote{An extra non-Abelian top-colour symmetry \cite{Hill:1991at} has been suggested in \cite{Dobrescu:1997nm} as an underlying theory. Extra vector-like quarks may also emerge in higher-dimensional extentions of the minimal top condensate model \cite{Dobrescu:1998dg}.}  results at a high energy scale $\Lambda$ in effective Nambu--Jona-Lasinio (NJL) type \cite{Nambu:1961tp} interactions involving the above quark fields:
\begin{eqnarray}
{\cal L}&=&{\cal L}_{kin}+\frac{g^2(\Lambda)\cos^2\theta}{\Lambda^2}\left(\bar Q_L\xi_R\right)\left(\bar \xi_R  Q_L\right)+\frac{g^2(\Lambda)\sin^2\theta}{\Lambda^2}\left(\bar Q_Lt_R
\right)\left(\bar t_R  Q_L\right) \nonumber \\
&+&\frac{g^2(\Lambda)\sin 2\theta}{2\Lambda^2}
\left[\left(\bar Q_L t_R\right) \left(\bar \xi_R  Q_L\right)+\text{h.c.}\right]-M_{\xi}(\Lambda)\left[\bar \xi_L\xi_R+\text{h.c.}\right]~,
\label{2}
\end{eqnarray}
where ${\cal L}_{kin}$ contains gauge invariant kinetic terms for fermionic fields (\ref{1}) and $g(\Lambda)$ and $\theta$ are parameters that define an overall and relative strength of 4-fermion interactions at the ultraviolet cut-off scale $\Lambda$. 
$M_{\xi}(\Lambda)$ is a running mass parameter, also defined at $\Lambda$. 

Next we assume that the underlying theory exhibits spontaneously broken scale invariance. In the low-energy theory (\ref{2}) it is non-linearly realised through the dilaton field $\chi$. Corresponding modification of the low-energy Lagrangian is achieved by the following replacements in Eq. (\ref{2}): 
\begin{equation}
\Lambda\to\frac{\Lambda}{f}\chi=\chi~,~M_{\xi}(\Lambda)\to\frac{M_{\xi}(\chi)}{f}\chi\equiv y_\xi(\chi)\chi~,
\label{3}
\end{equation}
where the dilaton decay constant $f$ is taken to be equal to $\Lambda$\footnote{Note that, if one in addition considers dilaton-gravity coupling and generation of the Planck mass, $f$ must be appropriately rescaled.}. The resulting scale invariant NJL Lagrangian can be rewritten in an equivalent form 
\begin{equation}
{\cal L}={\cal L}_{kin}-\left[y_{\xi}\bar \xi_L\xi_R \chi+g\bar Q_L\left( \cos\theta \xi_R+\sin\theta t_R\right)\Phi +\text{h.c.}\right]-\chi^2\left(\Phi^{\dagger}\Phi\right), 
\label{4}
\end{equation}
by introducing an auxiliary (non-dynamical at the scale $\Lambda$) electroweak doublet scalar field 
\begin{equation}
\Phi=-\frac{g}{\chi^2}\left( \cos\theta \bar \xi_R+\sin\theta \bar t_R\right)Q_L~.  
\label{5}
\end{equation}
${\cal L}_{kin}$ in (\ref{4}) now also includes the dilaton kinetic term, $1/2 (\partial_{\mu}\chi)^2$.  

Following \cite{Dobrescu:1997nm}, we assume that $\sin\theta \ll 1$. Hence, $\Phi$ is a composite field, given by $\left( \bar \xi_RQ_L\right)$ fermion bilinear predominantly. This non-dynamical field develops full dynamics at low energies and triggers the electroweak symmetry breaking. To see this we first integrate out modes of the fermionic fields in Eq. (\ref{1}) and those of dilaton with Euclidean momenta $(y_\xi \chi \leq) \mu < k_E<\Lambda (=\chi)$ in the functional integral. This results in the following 1-loop Higg-dilaton effective potential at scale $\mu_{\xi} =y_{\xi}\chi$ (the irrelevant operators with mass dimensions $d>4$ are omitted):     
\begin{eqnarray}
\begin{aligned}
&V=\sigma (\mu_{\xi}) \chi^2H^{\dagger}H+\lambda (\mu_{\xi}) \left(H^{\dagger}H\right)^2+\frac{\rho (\mu_{\xi})}{4}\chi^4~,  \\
&H:=\sqrt{Z_{\Phi}}\Phi,~~Z_{\Phi}(\mu_{\xi})=-\frac{3g^2}{8\pi^2}\ln |y_{\xi}|~,\\
&\sigma(\mu_{\xi})=
-\frac{8\pi^2}{3\ln |y_{\xi}|}\frac{\sigma}{g^2}\left(1+\frac{1-y_{\xi}^2}{16\pi^2}\right)+\frac{1-y_\xi^2}{\ln|y_\xi|}, \\
&\lambda(\mu_{\xi})=-\frac{8\pi^2}{3\ln |y_{\xi}|}\left(1-\frac{\sigma^2}{3g^4}\right)~,  \\
&\rho(\mu_{\xi})=-\frac{3y_{\xi}^2}{8\pi^2}\left(1-y_{\xi}^2+y_{\xi}^2\ln |y_{\xi}|\right)~.
\end{aligned}
\label{6}
\end{eqnarray}
Here $H$ denotes the electroweak doublet Higgs fields with a canonical kinetic term which is induced at low energies. Note that except of implicit scale dependence of dimensionless couplings in (\ref{6}), which is in turn governed by the vacuum expectation value of the dilaton field, $\langle\chi\rangle=\Lambda$, the theory exhibits classical scale invariance. In fact, the potential (\ref{6}) is similar to the one considered in our previous paper \cite{Kobakhidze:2017eml}, but is defined at the scale of the vectorlike fermion mass, $M_{\xi}=y_{\xi}\langle\chi\rangle < \Lambda$, rather than at the cut-off scale.  At the cut-off scale $\Lambda$ we instead have the compositeness condition, $Z_{\Phi}(\mu=\Lambda)=0$, which implies in particular that the renormalised top-Yukawa coupling developes Landau pole there, $y_t=g\sin\theta /Z_{\Phi}\stackrel{\mu\to\Lambda}{\longrightarrow}\infty$.  The top-quark mass then is determined by the infrared quasi fixed-point solution of renormalisation group equations (RGEs) for $y_t$ \cite{Pendleton:1980as} times an extra parameter $\sin\theta$, which can be adjusted to obtain experimental value of the top mass. Another important thing to note is that quantum fluctuations of dilaton also contribute to the Higgs self-coupling, such that $\lambda (\mu_{\xi})/y_t^2(\mu_{\xi})=\left(1-\frac{\sigma^2}{3g^4}\right)/\sin ^2\theta$, rather than the canonical value $1/\sin^2\theta$ of the original top condensate see-saw model. This allows us to adjust $\sigma$ and $g$ couplings such that the  experimental value of the Higgs mass can also be obtained (see below).

\section{Scalar masses}

Let us now turn to the minimization of the potential (\ref{6}). This leads to the following relation:
\begin{equation}
\frac{v^2}{\Lambda^2}=-\frac{\sigma(M_{\xi})}{\lambda (M_{\xi})}~, 
\label{7}
\end{equation}
which sets the hierarchy between the vaccum expectation values of the Higgs,  $\langle H \rangle=(0,v/\sqrt{2})^{\rm T}$ and dilaton, $\langle \chi \rangle=\Lambda$, fields, through the hierarchy of the dimensionless couplings. We would like to stress that this hierarchy is technically natural, since the coupling $\sigma$ is not generated quantum mechanically once is set to zero. While the ratio of scales is defined by Eq. (\ref{7}), the overall scale is defined through the dimensional transmutation, which through another minimization relation:
\begin{equation}
\sigma^2(M_{\xi})=\lambda (M_{\xi})\rho (M_{\xi})~,  
\label{8}
\end{equation}     

Finally, in order to comply with cosmological observations we impose a phenomenological condition of vacuum energy to be vanishingly small, $V(v,\Lambda)\simeq 0$. In scale invariant theories this implies an additional condition on dimensionless parameters \cite{Foot:2010et}, which in our case reads as: 
\begin{equation}
\frac{\beta_{\sigma}(M_{\xi})}{\sigma (M_{\xi})}\simeq \frac{\beta_{\lambda}(M_{\xi})}{2\lambda(M_{\xi})}+\frac{\beta_{\rho}(M_{\xi})}{2\rho(M_{\xi})},  
\label{9}
\end{equation}    
where, $\beta_{\sigma, \lambda,\rho}$ are RGE beta-functions for $\sigma, \lambda,\rho$ couplings, respectively. The later relation is the usual fine-tuning of cosmological constant. We have nothing new to say about this tuning here.  

Taking Eqs. (\ref{7},\ref{8},\ref{9}) into account, we compute the scalar mass squared matix in the $(h,\chi)$ basis, where $h$ is a neutral scalar component of the electroweak doublet $H$: 
\begin{eqnarray}
\hat {\rm M}^2(M_{\xi})=\left(\begin{tabular}{ll}
$ 2\lambda v^2~,$  & $2\sigma \left[1+\frac{\beta_{\sigma}}{2\sigma}-\frac{\beta_{\lambda}}{2\lambda}\right]\Lambda v$\\ 
$2\sigma \left[1+\frac{\beta_{\sigma}}{2\sigma}-\frac{\beta_{\lambda}}{\lambda}\right]\Lambda v~,$ & $2\rho \left[
1+\frac{\beta_{\sigma}}{\sigma}-\frac{\beta_{\lambda}}{\lambda} 
-\frac{\beta'_{\sigma}}{4\sigma}+\frac{\beta'_{\rho}}{8\rho}+\frac{\beta'_{\lambda}}{8\lambda}
\right]\Lambda^2$ \\ 
\end{tabular}\right) ,  
\label{10}
\end{eqnarray} 
 where the entries are running mass parameter evaluated at the scale $M_{\xi}$. As a consistency check we notice that the above matrix is degenerate in the conformal limit, i.e., when all the beta-functions (and their derivatives, $\beta'\equiv d\beta/d\ln\mu$) vanish. In this limit dilaton is the true Goldstone boson of spontaneously broken scale invariance and thus is massless. One also notices that the determinant of the above matrix is proportional to $\beta^2_{\sigma,\lambda}, \beta'_{\sigma, \lambda,\rho}$. This follows from the assumed fine-tuning of the cosmological constant, ref. Eq. (\ref{9}). Since $\beta^2_{\sigma,\lambda}, \beta'_{\sigma, \lambda,\rho}\sim {\cal O}(\hbar^2)$, we conclude that the dilaton mass is generated through the scale anomaly at 2-loop order. This is a generic feature of scale invariant models, and has been observed also in our previous works \cite{Foot:2010et}, \cite{Kobakhidze:2017eml}.  

Assuming the hierarchy of scales in Eq. (\ref{7}), we obtain that the dilaton and the neutral Higgs states have very small mixing, 
\begin{equation}
\sin 2\beta \simeq -\frac{v}{\Lambda}~,
\label{11}
\end{equation}  
and the mass of the Higgs boson is given essntially by the same formulae as in the Standard Model: 
\begin{equation}
m_h^2(M_{\xi})\simeq 2\lambda (M_{\xi})v^2~.
\label{12}
\end{equation} 
The effective running mass for dilaton can also be computed straightforwardly:
\begin{equation}
m_{\chi}^2(M_{\xi})=-2\sigma(M_{\xi})\left.\left[-\frac{1}{4}\left(\frac{\beta_{\sigma}}{\sigma}-\frac{\beta_{\lambda}}{\lambda}\right)^2-\frac{\beta'_{\sigma}}{4\sigma}+\frac{\beta'_{\rho}}{8\rho}+\frac{\beta'_{\lambda}}{8\lambda}\right]\right |_{\mu=M_{\xi}}v^2~. 
\label{13}
\end{equation} 
We observe that the dilaton mass is hierarchically smaller than the Higgs mass, $m_{\chi}/m_{h}\propto v/\Lambda$, being in addition suppressed by a 2-loop factor, see Eq. (\ref{13}). Thus, for large $M_{\xi}\gg v$, the current model is phenomenologically indistinguishable from the scale invariant Standard Model proposed in   
\cite{Kobakhidze:2017eml}. 
 
\section{Conclusions}
In this paper we have constructed a phenomenologically viable top condensate see-saw model with spontaneously broken scale invariance. The hidden scale invariance is  implemented at low energies by introducing the dilaton field. The compositness scale is generated quantum mechanically via the mechanism of dimensional transmutation and the technically natural hierarchy between the electroweak scale and high energy scales  is maintaned. The Higgs boson is a composite state of left-handed top quark and (predominantly) of right-handed component of massive vectorlike anti-quark and essentially indistinguishable from the Standard Model Higgs boson. The critical prediction of this model is the very light dilaton, detection of which would be an important hint of the relevance of scale invariance in the electroweak symmetry breaking.      

\paragraph{Acknowledgements.}

We would like to thank Lewis Tunstall for the collaboration at earlier
stages of this project. The work was supported in part by the Australian
Research Council.

\end{document}